\documentclass[a4paper,fleqn,10pt]{article}
\usepackage{amsmath}
\usepackage{amssymb}
\usepackage{array}
\usepackage{calc}
\usepackage{longtable}
\usepackage{multirow,booktabs}
\usepackage{relsize}
\usepackage{pstricks}
\usepackage{graphicx}
\usepackage{xspace}
\usepackage{units}

\numberwithin{equation}{section}
\usepackage{mciteplus}
\usepackage[a4paper,pdfborder={0 0 0}]{hyperref}
\usepackage[format=hang,labelfont=bf,hypcap=true]{caption}
\usepackage{subcaption}
\usepackage{sectsty}
\allsectionsfont{\sffamily}
\subsubsectionfont{\sffamily}
\makeatletter
\DeclareRobustCommand*{\bfseries}{%
  \not@math@alphabet\bfseries\mathbf
  \fontseries\bfdefault\selectfont
  \boldmath
}
\makeatother
\let\spreprint\empty
\newcommand{\preprint}[1]{\def\spreprint{\protect#1}}
\let\sinstitute\empty
\newcommand{\institute}[1]{\def\sinstitute{\protect#1}}
\makeatletter
\renewcommand{\maketitle}{\begingroup
  \null\thispagestyle{empty}%
    \ifx\spreprint\empty
      \vskip 5ex
    \else
      \flushright\large\spreprint\vskip 2ex
    \fi
    \vskip 5ex
    \flushleft
      {\sffamily\bfseries\huge\@title}\vskip 6ex
      \@author\vskip 2ex
      \ifx\sinstitute\empty
      \else
        {\small\sinstitute}
      \fi
    \vskip 5ex
  \endgroup
}
\makeatother
\renewenvironment{abstract}{\begin{center}
  {\large\sffamily\bfseries Abstract: }
  \begin{minipage}[t]{0.75\textwidth}
}{\end{minipage}\end{center}\vskip 10ex}

\numberwithin{equation}{section}
\DeclareRobustCommand{\plusplus}{\raisebox{0.2ex}{\smaller++}}

\newcommand{\done}{{\mathrm d}}

\newcommand{\bea}{\begin{align}}
\newcommand{\eea}{\end{align}}
\newcommand{\beq}{\begin{equation}}
\newcommand{\eeq}{\end{equation}}

\newcommand{\bs}{\begin{split}}
\newcommand{\es}{\end{split}}

\newcommand{\bi}{\begin{itemize}}
\newcommand{\ei}{\end{itemize}}
\newcommand{\bc}{\begin{center}}
\newcommand{\ec}{\end{center}}
\newcommand{\bac}{\begin{array}{c}}
\newcommand{\bacc}{\begin{array}{cc}}
\newcommand{\ea}{\end{array}}

\def\spa#1.#2{\langle#1\,#2\rangle}
\def\spb#1.#2{[#1\,#2]}

\newcommand{\kt}{\ensuremath{{\bf k}_\perp}\xspace}

\newcommand{\Dipole}{\ensuremath{{\mathcal{D}}}}

\newcommand{\IFDipole}[3]{\ensuremath{{\mathcal{D}^{#1#2}_{#3}}}}
\newcommand{\IIDipole}[3]{\ensuremath{{\mathcal{D}^{#1#2;#3}}}}

\newcommand{\Sub}{\ensuremath{{\mathcal{S}}}}

\newcommand{\Born}{\ensuremath{{\mathcal{B}}}}

\newcommand{\Real}{\ensuremath{{\mathcal{R}}}}
\newcommand{\Subtraction}{\ensuremath{{\mathcal{S}}}}

\newcommand{\Virtual}{\ensuremath{{\mathcal{V}}}}

\newcommand{\OnePS}{\ensuremath{\Phi_{1}}}
\newcommand{\PS}[1]{\ensuremath{\Phi_{#1}}}

\newcommand{\sla}[1]{\ensuremath{{#1\kern-0.45em/}}}

\newcommand\LHC{L\protect\scalebox{0.8}{HC}\xspace}

\newcommand{\MCatNLO}{M\protect\scalebox{0.8}{C}@N\protect\scalebox{0.8}{LO}\xspace}

\newcommand{\Dire}{D\protect\scalebox{0.8}{IRE}\xspace}

\newcommand{\OpenLoops}{O\protect\scalebox{0.8}{PEN}L\protect\scalebox{0.8}{OOPS}\xspace}

\newcommand{\Sherpa}{S\protect\scalebox{0.8}{HERPA}\xspace}

\newcommand{\Amegic}{A\protect\scalebox{0.8}{MEGIC\plusplus}\xspace}
\newcommand{\CSS}{C\protect\scalebox{0.8}{SS}\xspace}

\newcommand{\Rivet}{R\protect\scalebox{0.8}{IVET}\xspace}

\newcommand{\Fastjet}{F\protect\scalebox{0.8}{AST}J\protect\scalebox{0.8}{ET}\xspace}

\hypersetup{
  pdfauthor={}
  pdftitle={}
}
\preprint{IPPP/17/105\\DCPT/17/210\\MCNet/17/24}
\author{Frank Krauss$^1$, Davide Napoletano$^{1,2}$}
\title{Towards a fully massive five-flavour scheme}
\institute{
  $^1$ Institute for Particle Physics Phenomenology,
  Durham University, Durham DH1 3LE, UK\\
  $^2$ IPhT, CEA Saclay, CNRS UMR 3681, F-91191, Gif-Sur-Yvette, France}
\begin{document}
\maketitle
\begin{abstract}
  In this work we explore first necessary steps to contruct a fully massive
  version of a variable flavour number scheme. In particular we focus, as an
  example, on an extension of the five-flavour scheme, where instead of
  neglecting explicit initial state quark mass effects, we retain all massive
  dependence, while keeping the resummation properties of the massless
  five-flavour scheme.  We name this scheme five-flavour-massive (5FMS) scheme.
  Apart from consistently modified parton distribution functions, we provide
  all the ingredients that are needed to implement this scheme at \MCatNLO
  accuracy, in a Monte Carlo event generator.  As proof of concept we
  implement this scheme in \Sherpa, and perfom a comparison of the new scheme
  with traditional ones for the simple process of scalar particle production
  in bottom quark fusion.
\end{abstract}
\section{Introduction}
\label{sec:mass_sub}
Processes with heavy quarks (bottom or charm) in the initial state present an
interesting challenge for theoretical predictions at the \LHC and other hadron
collider experiments.  First, the finite quark masses introduce another
scale to the process, which may or may not prove to be relevant for different
observables and different processes.  In addition, a decision has to be made
in how far heavy quarks can act as incident partons -- due to their mass being
larger than the QCD scale parameter $m_Q\gg \Lambda_{\rm{QCD}}$ one could argue
that they are disallowed to have a parton distribution function (PDF), thereby
decoupling them from the QCD evolution in the initial state, described by the
DGLAP equations.  This leads to two complementary solutions.  On one hand,
heavy quarks $Q$ in the initial and final state may be treated on the same
footing as any other light quark, such as the $u$, $d$, or $s$ quark, by
ignoring their mass in the evaluation of the matrix elements.  In such a
picture the heavy quark acts as an active quark in the QCD evolution
equations, and consequently, possibly large collinear logarithms are resummed
to all orders into a $Q$ PDF.  On the other hand, for some processes and
observables the effects of the finite heavy quark mass, $m_Q$ become relevant
and in such cases these quarks must be treated as fully massive.
Traditionally, this immediately translates into the heavy quarks only appearing
as final state particles. 

This dichotomy is most pronounced for the case of $b$ quarks, due to their
mass $m_b\approx 4.5$ GeV being larger than the charm mass by a factor of
about 3.  It gives rise to ongoing comparisons of calculations of the same
processes and observables in the {\it five-flavour} and {\it four-flavour}
schemes.  Here, the former refers to a consistently massless treatment of
the $b$ quark, which can therefore be found in both initial and final states,
while the latter treats the $b$ quarks as massive and allows them to be in
the final state only.  For a recent example focusing on the production of $Z$
or Higgs bosons in association with $b$ quarks at the \LHC,
cf.~\cite{Krauss:2016orf}.  There, a slight preference for five-flavour
scheme calculations in a multijet merging approach has been found.  Broadly
speaking, for a wide range of kinematical observables such as the $p_\perp$
spectrum of jets or gauge or Higgs bosons away from small momenta, this is
in agreement with other, similar studies~
\cite{Chatrchyan:2013zja,Aad:2014dvb,Khachatryan:2016iob}.
A preferable solution, would be to perform a matching between these two scheme,
see for examples~\cite{Aivazis:1993pi,Aivazis:1993kh,Thorne:1997ga,Thorne:1997uu,
  Cacciari:1998it,Forte:2010ta,Forte:2015hba,Bonvini:2015pxa,Forte:2016sja,
  Bonvini:2016fgf}.
However, so far, these schemes have only been worked out for inclusive
enough observables, and are not yet suitable for a Monte Carlo implementation.

This finding motivates to extend the five-flavour scheme to allow massive
particles in the initial state.  In this paper, we present the necessary
ingredients for {\it next-to-leading} order calculations with massive initial
state partons, including these mass effects in initial state parton shower.
We refer to the extended scheme as {\it five-flavour-massive} scheme, or 5FMS
in short.  This scheme thus has massive $b$-quarks that contribute both to
the running of the coupling constant and to the evolution of PDFs.

There are a number of obstacles to this goal:
\begin{enumerate}
\item in order to calculate cross sections at next-to leading order accuracy
  in the strong coupling, a scheme to identify, isolate and subtract infrared
  divergences is yet to be worked out in full detail.  In particular, we
  follow the logic of the Catani-Seymour subtraction formalism, which was
  first presented for massless partons in~\cite{Catani:1996vz}, and later
  extended to massive fermions in QED in~\cite{Dittmaier:1999mb}, to massive
  final state QCD partons in~\cite{Catani:2002hc}, and to massive initial
  state quarks for initial-final dipoles in~\cite{Kotko:2012ui}.  We work out
  phase-space mappings and differential and integrated splitting kernels
  for the emission of a gluon off a massive quark line in the initial state,
  with an initial state spectator.  

  The treatment of massive initial state particles in QED has already been
  discussed in~\cite{Dittmaier:1999mb}.  However, in contrast to the results
  there, obtained in $D=4$ dimensions with a massive photon with $m_\gamma$ as
  infrared regulator, we consistently work in $D=4-2\varepsilon$ dimensions
  with a massless gluon.  Of course, expressions can be mapped onto each other
  by suitably replacing:
  \begin{equation}
    e^2\,Q_a\,\sigma_a\,Q_b\,\sigma_b \leftrightarrow
    4\,\pi\,\alpha_S\,\mu^{2\varepsilon}\,{\bf T}_a\cdot{\bf T}_b\,,
  \end{equation}
  and, working in the $\overline{\text{MS}}$ scheme,
  \begin{equation}
    \log m_{\gamma}^2 \leftrightarrow
    \frac{1}{\varepsilon} + \log 4\,\pi\,\mu_R^2 + \cal{O}(\varepsilon)\,.
  \end{equation}

\item standard five-flavour PDFs introduce massive quarks purely perturbatively,
  through gluon splitting within the evolution.  In so doing, special care is
  devoted to the treatment of threshold effects due to the finite masses,
  resulting in {\it variable-flavour number schemes}, such as the ones detailed
  in~\cite{Aivazis:1993pi,Aivazis:1993kh,Thorne:1997ga,Thorne:1997uu,
    Forte:2010ta}.  However, all these schemes treat mass effects only through
  thresholds and usually ignore other kinematical effects.  We modify standard
  PDFs through a number of plausible choices detailed below to obtain some
  handle on the size of such effects.  However, a full and comprehensive study
  of mass effects in PDFs is beyond the scope of this publication.  This is
  also true for more conceptual questions in how far such mass effects must
  be treated as process-dependent corrections, similar to higher-twist effects.
  While we acknowledge that these may be important considerations, 
  we leave the detailed study of these effects in hadron-hadron collision
  for a separate work.
\end{enumerate}
An additional problem, as established
in~\cite{Catani:1985xt,Catani:1987xy,Doria:1980ak,DiLieto:1980nkq,Catani:2002hc},
is that starting at NNLO there are non-cancelling infrared divergent contributions
that are proportional to the initial state quark mass. This renders the scheme
presented in this paper only valid up to NLO accuracy.

Lastly, approaches that use a finite heavy quark mass in the parton shower
have been studied in literature\cite{Nagy:2007ty,Nagy:2014oqa,Nagy:2014mqa}.
Although a comparison is certainly interesting, we leave this to future
studies.

The outline of this work is the following. In the next section we very briefly
summarize the Catani--Seymour subtraction procedure.  There, we also report
the ingredients needed to extend this method to include massive initial state
quarks.  In section~\ref{sec:mcatnlo} we present the relevant modifications
for the matching of the parton shower and next-to-leading order matrix elements.
We provide the discussion of results in section~\ref{sec:results} where 
we show explicit results for the
production of a scalar boson, $A$, in bottom-quark fusion for various
combinations of ($m_A,\tan\beta$) both at fixed-order and matched to the
parton shower.  There we also compare our results with the \Dire parton
shower, which includes (collinear) NLO corrections to the DGLAP
equation~\cite{Hoche:2017iem,Hoche:2017hno}.

\section{Catani--Seymour subtraction for massive initial states}
\label{sec:sub}
\subsection{Nomenclature}
The differential leading-order cross section (LO) for a hard scattering
process with $N$ particles in the final state is given by
\begin{equation}
  \label{eq:lodsigma}
        \done\hat{\sigma}_{ab} =
        \done\PS{N}(p_1,\dots,p_N)\,
        \Born_N(p_a,p_b;p_1,\dots,p_N)\,
        F_J^{(N)}\left(p_1,\dots,p_N;p_a,p_b\right)\,,
\end{equation}
where $\Born$ denotes the Born matrix element squared and the differential
phase space element $\done\PS{N}$ implicitly contains the incoming flux of
the incident particles, and parton distribution functions, where applicable.
Later, where they matter, we will make these factors explicit.  The measurement
function, $F_J^{(N)}$, guarantees that the $N$-jet final state is well defined
at Born level and for Born kinematics.  In a similar fashion, and suppressing
the obvious four-vectors as arguments, the cross section at next-to leading
order (NLO) cross section is given by
\begin{equation}
  \label{eq:signlo}
  \sigma_N^{\text{NLO}} =
  \int\done \PS{N}\left[\Born_N\,+\,\Virtual_N\right]\, F_J^{(N)}+\,
  \int \done \PS{N+1} \Real_N\, F_J^{(N+1)},
\end{equation}
where the additional terms $\Virtual_N$ and $\Real_N$ signify the virtual
and real corrections to the original Born term.  They of course relate to
final states with $N$ and $N+1$ particles, respectively, as indicated by the
phase space elements. The measurement function must satisfy
\begin{equation}
  F_J^{(N+1)}\left(p_1,\dots,k,\dots,p_N;p_a,p_b\right)\rightarrow
  F_J^{(N)}\left(p_1,\dots,p_N;p_a,p_b\right)\quad \text{if}
  \quad k\cdot\{ p_I\}_{I=i,a,b}\rightarrow 0\,\lor \,|k|\rightarrow 0 
\end{equation}
to ensure a meaningful cross section definition at NLO accuracy.  In later
parts of this publication we will assume that this function is implicitly
included.

The soft and collinear divergences related to the emission of the additional
particle in $\Real_N$ are cancelled by similar structures in the virtual
part $\Virtual_N$, but in order to facilitate this cancellation the poles
in both must first be isolated and dealt with.  Subtraction methods, such as
Catani--Seymour subtraction, make use of the universal property of QCD
amplitudes in the soft and collinear limits, where the corresponding divergent
poles factorize.  This allows the construction of process-independent
subtraction terms $\Subtraction_N(\PS{N}\otimes\PS{1})$, such that the
first term on the right hand side of
\begin{multline}
  \label{eq:subk}
  \int \done \PS{N+1} \Real_N(\PS{N+1}) = \int \done \PS{N+1}
  \left[\Real_N(\PS{N+1})-\Subtraction_N(\PS{N}\otimes\PS{1})
    \vphantom{\frac12}\right]F_J^{(N+1)}  \\+
  \int \left[\done \PS{N} \otimes \done\PS{1}\vphantom{\frac12}\right]
  \Subtraction(\PS{N}\otimes{\Phi}_{1})\,F_J^{(N)}\hspace{1cm}
\end{multline}
is finite.  Assuming that infrared divergences stem from regions in phase
space where the momentum $p_k$ of a particle $k$ becomes soft or collinear to
a particle $i$ with momentum $p_i$, the degree of infrared divergence can
be parametrized by a small $\lambda\rightarrow 0$ such that
$|p_k|\sim\lambda$ or $p_i\cdot p_k\sim\lambda$.  In these divergent phase
space regions, the difference in the first term on the r.h.s.\ of the equation
above behaves as
\begin{equation}
  \label{eq:diplim}
  \lim_{\lambda\rightarrow 0}
  \left[\vphantom{\frac 12}\Real_N(\PS{N+1})-
    \Subtraction_N(\PS{N}\otimes\PS{1})\right]_{p_k,p_i\cdot p_k\sim\lambda} =
  \mathcal{O}(\lambda^0)\,,
\end{equation}
{\it i.e.}\ all infrared poles have been cancelled.

In addition, the functions $\Subtraction(\PS{N}\otimes\PS{1})$ are
constructed in such a way that their integral over the extra emission
phase space -- the second term on the right hand side of Eq.(\ref{eq:subk}) --
can be calculated analytically in $D=4+2\varepsilon$ dimensions, with their
divergent parts giving rise to poles $1/\varepsilon^2$ and $1/\varepsilon$.
These poles are ultimately combined with the infrared poles from the loop
contributions to cancel exactly.

Combining Eqs.(\ref{eq:signlo}) and (\ref{eq:subk}) yields
\begin{equation}
  \label{eq:fsubk}
  \sigma_N^{\text{NLO}} =
  \int\done \PS{N}
  \left[\Born_N\,+\,\Virtual_N\,+\,\int\done\OnePS\,\Sub_N\right]\,F_J^{(N)}
  \,+\, \int \done \PS{N+1}
  \biggl[\Real_N-\Subtraction_N\biggr]\,F_J^{(N+1)}, 
\end{equation}
where each phase space integral by itself is infrared finite.

In Catani--Seymour subtraction, the terms $\Sub$ are formulated in terms of
dipoles, made from three particles, an emitter, a spectator and the emitted
particle $k$.  The subtraction term factorizes into a product of
process-independent emission terms and Born-like configurations, possibly
with parton flavours that differ from the original Born term.  These dipoles
$\Dipole$ are then classified by the splitter and spectator parton being
either in the initial (I) or final (F) state, as II, FI, IF, or FF -- the
emitted particle $k$ obviously always is in the final state.  The overall
subtraction term therefore reads
\begin{equation}
  \label{eq:dds}
  \Subtraction \equiv \sum_{i=\text{FF,FI,IF,II}}\,\Dipole_{i},
\end{equation}
where each dipole contribution is given by the sum of all possible
emitter--spectator pairs,
\begin{eqnarray}
  \label{eq:ddipoles}
  \begin{array}{lclclcl}
  \Dipole_{FF} & = & \sum\limits_{(i,j)\neq k}\,\Dipole_{ij,k}\,; &\quad&
  \Dipole_{FI} & = & \sum\limits_{(i,j)\neq k}\,\Dipole_{ij,k}\,+\,\Dipole_{ij}^a
  \\[5mm]
  \Dipole_{IF} & = & \sum\limits_{i\neq k}\,\IFDipole{a}{i}{k}\,;&\quad&
  \Dipole_{II} & = & \IIDipole{a}{k}{b} \,+ \,\IIDipole{b}{k}{a}\,.
\end{array}
\end{eqnarray}

In the context of this study we primarily focus on II configurations, which
can most conveniently be studied in quark--annihilation processes such as
$b\bar{b}\to H$ and similar.  For these processes,
\begin{equation}
  \Subtraction = \IIDipole{a}{k}{b}(p_1,\dots ,p_k,\dots,p_{N+1};p_a,p_b) \,
  + \,\IIDipole{b}{k}{a}(p_1,\dots ,p_k,\dots,p_{N+1};p_a,p_b) \,.
\end{equation}
The term $\IIDipole{a}{k}{b}$
\begin{equation}
  \label{eq:dipoles}
  \IIDipole{a}{k}{b}(p_1,\dots ,p_k,\dots,p_{N+1};p_a,p_b) =
  -\frac{ 1}{2\,x_{ab}\,p_a\cdot p_k}\frac{{\bf T}_a\cdot{\bf T}_b}{{\bf T}_a^2}
  \,{\bf V}^{ak,b}(p_a,p_b,p_k) 
  \otimes {\left|\widetilde{\mathcal{M}}_N
    \left(\widetilde{p}_1,\dots ,\widetilde{p}_{N+1};
    \widetilde{p}_a,p_b\right)\right|}^2
\end{equation}
represents one individual dipole contribution, where the emitter is the initial
state particle $a$ and the spectator is the other initial state particle $b$.
The matrix element $\widetilde{\mathcal{M}}$ emerges from the original
Born-level matrix element by taking into account that the emission of $k$
off parton $a$ might alter the flavour of the resulting parton $\tilde{a}$,
and it is evaluated at a kinematical configuration, where the modified
four-vectors $\tilde{p}_a$ and $\tilde{p}_i$ account for four-momentum
conservation by absorbing the recoil from emitting $p_k$.  The functions
$\frac{1}{2\,p_i\cdot p_j}{\bf V}^{ij,k}(p_i,p_k,p_j)$ are generically called
splitting kernels, or dipole splitting functions and reduce to the well-known
Altarelli-Parisi~\cite{Altarelli:1977zs}
splitting kernels $P_{ij}$ in the collinear limit.  Note that
the $\otimes$ symbol implies possible summation in color and helicity space.

\subsection{Massive II dipoles:
  initial state splitter with initial state spectator}
We now present the relevant modification to the described picture, due to the
inclusion of finite masses.  In the following we make extensive use of the
following kinematical quantities:
\begin{eqnarray}
  \begin{array}{lclclcl}
    s     & = &(p_a \,+\,p_b)^2\,; &&
    Q^2   & = & (p_a \,+\,p_b\,-\,p_k)^2 = s\,-2\,\,(p_a \,+\,p_b)\cdot p_k\,; 
    \\[5mm]
    x_{ab} & = & \displaystyle
    \frac{p_a\cdot p_b \,-\,p_a\cdot p_k \,-\,p_b\cdot p_k}
         {p_a\cdot p_b}\,; &&
    y_{a}  & = & \displaystyle\frac{p_a\cdot p_k}{p_a\cdot p_b}\,; 
    \\[5mm]
    s_{ab} & = & s\,-\,m_a^2\,-\,m_b^2\,; &&
    \lambda_{ab} & = & \lambda(s,m_a^2,m_b^2)\,,
  \end{array}
\end{eqnarray}
where
\begin{equation}
  \lambda(a,b,c)=a^2+b^2+c^2-2\,(ab+bc)-2\,ac\,.
\end{equation}

The only dipoles involving massive partons which exhibit infrared divergences
are those corresponding to the emission of a gluon into the final state.
Therefore, we only have to find a suitable expression for the term
${\bf V}^{ak,b}(p_a,p_b,p_k)$ for the case where a heavy initial quark $Q$
emits a gluon with momentum $p_k$.  In analogy to the treatment for massive
final state particles in~\cite{Catani:2002hc} this is given by
\begin{equation}
  \label{eq:diff_sub}
  {\bf V}^{Q_ag_k,b}(p_a,p_b,p_k)\,=\,8\,\pi\,\mu^{2\varepsilon}\alpha_s\,C_F\,
  \left[\frac{2}{1-x_{ab}}\,-(1\,+\,x_{ab})-\varepsilon(1\,-\,x_{ab})-
    \frac{x_{ab}\,m_a^2}
    {p_a\cdot p_k}\right]\,.
\end{equation}

In any other case --- $g\to Q\bar{Q}$ and $Q\to gQ$ --- there are no singular
contributions rendering the need for subtraction obsolete.  It is clear,
however, that the collinear divergences present for the very same splittings
in the massless case give rise to logarithmically enhanced terms of the form
$\log m_Q^2/\mu^2$, where $\mu$ is some large scale related to the dipole
kinematics.  While one may be tempted to use subtraction terms to smooth
these structures and make them more amenable to numerical integration, we have
tested explicitly that they do not pose any problem for processes at 
\LHC energies and masses down to 1 GeV. In any case detailed expressions
in the QED case in four space-time dimensions can be found in~\cite{Dittmaier:2008md}.

Coming back to the case of gluon emissions off a heavy quark line, we set
the subtraction term to zero for $x_{ab}<\alpha$, the kinematical lower bound,
\begin{equation}
  \alpha = \frac{2\,m_a^2\,m_b^2}{s_{ab}}\,.
\end{equation}
In a next step we need to define the phase space map, connecting the
original momenta $\{p_i\}$ of the real emission configuration to the modified
momenta $\{\tilde{p}_i\}$ for the reduced matrix element in the subtraction
term.  This map has to preserve mass-shell conditions, and in particular
$\tilde{p}_a^2=p_a^2 = m_a^2$, and it is also customary to keep the spectator
momentum fixed.  As a consequence of these conditions, all other final state
momenta $\tilde{p}_j$ and their total momentum $\tilde{Q}=\sum_j\tilde{p}_j$
absorb the recoil in the reduced matrix element.

The transformations are given by,
\begin{eqnarray}
  \label{eq:trans}
  \widetilde{p}_a^\mu & = &
  \sqrt{\frac{\lambda\left(Q^2,m_a^2,m_b^2\right)}{\lambda_{ab}}}\,
  \left(p_a^\mu-\frac{s_{ab}}{2\,m_b^2}p_b^\mu\right)\,+\,
  \frac{Q^2-m_a^2-m_b^2}{2\,m_b^2}p_b^\mu\,;\nonumber\\
  \widetilde{Q}^\mu & = & \widetilde{p}_a^\mu + p_b^\mu\,;\nonumber\\
  \left. \widetilde{p}_i^\mu\right|_{i\neq k} & = &
  \Lambda_{\nu}^\mu \, \left. p_i^\nu\right|_{i\neq k}\,,
\end{eqnarray}
where the Lorentz transformation $\Lambda_{\nu}^\mu$ is given by
\begin{equation}
  \Lambda_{\nu}^\mu = g_{\nu}^\mu
  - \frac{(Q+\widetilde{Q})^\mu
    (Q+\widetilde{Q})_\nu}{Q^2+Q\cdot\widetilde{Q}}
  + \frac{2\,\widetilde{Q}^\mu{Q}_\nu}{Q^2}
\end{equation}
and applied to all final state particles apart from $k$, including
colourless ones.

It is straightforward to check that these relations fulfil the mass-shell
conditions, such that $\widetilde{p}_a^2=m_a^2$ and $\widetilde{Q}^2=Q^2$,
and that they possess the right infra-red and collinear asymptotic limits.

\subsection{Phase space}
The phase space for the real emission correction factorises into a Born-level
part and a one-particle phase space integral,
\begin{equation}
  \int \done\PS{N+1}(p_k,Q;p_a+p_b) = \int\limits_{\alpha}^1\done x
  \int\done\PS{N}(\widetilde{Q}(x);\widetilde{p}_a(x)+p_b)
  \int\left[\done^{d-1}p_k(s,x,y_{a})\right]\,,
\end{equation}
where $x$-dependent momenta can be obtained from $\widetilde{p}_a$
and $\widetilde{Q}$ upon replacing $Q^2\rightarrow s_{ab}\,x +m_a^2+m_b^2$.
The extra particle phase space reads
\begin{equation}
  \label{eq:phase-space}
  \int\left[{\rm d}^{d-1}p_k(s,x,y_{a})\right] =
  \frac{1}{16\,\pi^2}\frac{(4\pi)^\varepsilon}{\Gamma(1-\varepsilon)}
  {\left(\frac{s_{ab}}{\sqrt{\lambda_{ab}}}\right)}^{1-2\varepsilon}
  {\left(1-x\right)}^{1-2\varepsilon}\,
    s_{ab}\, s^{-\varepsilon} \int_{v_2}^{v_1}{\rm d}v
  {\left[\left(v_1-v\right)\left(v-v_2\right)\right]}^{-\varepsilon}\,,
\end{equation}
where, for convenience, we define  $ v = y_{a}/\,(1-x)$, and
\begin{equation}
  v_{1,2} = \frac{s_{ab}\,+\,2\,m_a^2 \,\mp\,\sqrt{\lambda_{ab}}}{2\,s}\,.
\end{equation}

The integrated splitting function $\widetilde{\Virtual_N}^{a,b}$ is given by
\begin{equation}
  \frac{\alpha_s\, C_F}{2\,\pi}
  \widetilde{\Virtual_N}^{a,b}(x;\varepsilon) =
  \int\left[\done^{d-1}p_k(s,x,y_{a})\right]
  \frac{1}{2p_a\cdot p_k} {\bf V}^{ak,b}\,.
\end{equation}
and can be decomposed according into an end-point contribution
$\Virtual_N^{a,b}(\varepsilon)$
containing the $1/\varepsilon$ pole and a finite part $K^{a,b}(x)$
\begin{equation}
  \label{eq:split_delta_k_term}
  \widetilde{\Virtual_N}^{a,b}(x;\varepsilon) =
  \delta(1-x)\,\Virtual_N^{a,b}(\varepsilon)
  +\left[ K^{a,b}(x)\right]_+.
\end{equation}
The individual pieces read
\begin{eqnarray}
  \label{eq:intsub}
  \Virtual_N^{a,b}(\varepsilon) &=&\;\;
  \frac{1}{\varepsilon}\,
  \left(1 + \frac{s_{ab}}{\sqrt{\lambda_{ab}}\log\beta_0}\right)
  + \log{\left(\frac{\mu^2\,s}{s_{ab}^2}\right)} +\frac{3}{2}
  \nonumber\\
  && +\frac{s_{ab}}{\sqrt{\lambda_{ab}}}
  \left\{\left[\log{\left(\frac{\mu^2\,\lambda_{ab}}{m_a^2s_{ab}^2}\right)}
      + \left(\frac{1}{2} - \frac{2\,m_a^2}{s_{ab}}\right)\right]\log\beta_0 +
  2\,{\rm Li}_2(\beta_0) +\frac{1}{2}\log^2\beta_0-\frac{\pi^2}{3}\right\}
  \nonumber\\
  K^{a,b}(x) &=& -\frac{s_{ab}}{\sqrt{\lambda_{ab}}}\left(\frac{1\,+\,x^2}
  {1-x}\right)\log\beta_0\,-2\,x\,,
\end{eqnarray}
where
\begin{equation}
  \beta_0 =
  \frac{s_{ab}\,+\,2\,m_a^2 \,-\,
    \sqrt{\lambda_{ab}}}{s_{ab}\,+\,2\,m_a^2 \,+\,\sqrt{\lambda_{ab}}}\,.
\end{equation}
Note that due to the absence of a collinear divergence -- which is shielded
by the finite quark mass -- there are no terms $\propto 1/\varepsilon^2$.

The contribution to the partonic differential cross section is thus given by
\begin{eqnarray}
  \label{eq:part_I}
  \int\done\OnePS\,\Sub &=&
  \frac{\alpha_s\, C_F}{2\,\pi}
  \left\{
  \left[\Virtual_N^{a,b}(\varepsilon)-\int\limits_{0}^\alpha\done x\,K^{a,b}(x)\right]
  \,\done\PS{N}\Born(s_{ab}) \right.\nonumber\\
  &&\left.\hspace*{10mm}
   +  \int\limits_{\alpha}^1\done x\,K^{a,b}(x)
    \left[
    \frac{\phi(x\,s_{ab})}{x\,\phi(s_{ab})} \done\PS{N} \,\Born_N(x\,s_{ab}) -
    \done\PS{N} \,\Born_N(s_{ab})\right]
  \right\}\, ,
\end{eqnarray}
where we make explicit the dependence on the is the initial state flux, $\phi$.

To embed this into the calculation of cross section at hadron colliders, the
PDFs must be added.  Parametrizing the incoming hadron and parton momenta as
\begin{eqnarray}
  P_{1,2}^\mu & = & \frac{\sqrt{S}}{2}\,\left(1,0,0,\pm\,1\right)\,,\nonumber\\
  p_{a,b}^\mu & = & \eta_{1,2}\,P_{1,2}\,+,\frac{m_{a,b}^2}{\eta_{1,2}\,S}\,P_{2,1}\,
\end{eqnarray}
yields the allowed intervals for the light-cone momentum fractions
\begin{equation}
  \eta_{1,2}\,\in\,
  \left[\frac{1}{2}\left(1\,-\,\sqrt{1-\frac{4\,m_a^2m_b^2}{S^2}}\right),\,
    \frac{1}{2}\left(1\,+\,\sqrt{1-\frac{4\,m_a^2m_b^2}{S^2}}\right)\right]\,.
\end{equation}
Making explicit the flux, $\phi(s_{ab}) = 4\,\sqrt{\lambda(s,m_a^2,m_b^2)}$,
the integral over the incoming light-cone momenta, and the parton distribution
functions, the integrated splitting function, corresponding to the purely
partonic expression in Eq.~(\ref{eq:part_I}), reads
\begin{eqnarray}
  \label{eq:had_I}
  \mathcal{I} &=& \frac{\alpha_s\, C_F}{2\,\pi} \int \done\eta_1\,\done\eta_2\,
  f_a(\eta_1)\,f_b(\eta_2)
  \left\{
  \left[\Virtual_N^{a,b}(\varepsilon)-\int\limits_{0}^\alpha\done x\,K^{a,b}(x)\right]
  \,\done\PS{N}\Born(s_{ab}) \right.\nonumber\\
  && \left.\hspace*{4.5cm} +  \int\limits_\alpha^1\done x\,K^{a,b}(x)
    \left[
    \frac{\phi(x\,s_{ab})}{x\,\phi(s_{ab})} \done\PS{N} \,\Born_N(x\,s_{ab}) -
    \done\PS{N} \,\Born_N(s_{ab})\right]
  \right\}\,.\nonumber\\
\end{eqnarray}
In this form $\mathcal{I}$ is not very useful for direct implementation, because
the first term in the second line implies that the parton-level Born cross
section must be integrated over all values of $x$ in the interval $[\alpha,\,1]$.  
To remedy this, we need to rewrite this term, to disentangle the Born cross
section from the $x$-integration.  To fix this we define the following
variable transformation,
\begin{align}
  x\, s_{ab}(\eta_1,\eta_2) = s_{ab}(\eta^\prime_1(x),\eta^\prime_2)
  \;\;\;\mbox{\rm and}\;\;\;
  \eta^\prime_2 = \eta_2\,,
\end{align}
which defines a Jacobean, $J(\eta_1^\prime(x),\eta_2)$, and
\begin{equation}
  \done\eta_1\done\eta_2\done x =
  \done\eta_1^\prime\done\eta_2\done x J(\eta_1^\prime(x),\eta_2)\,,
\end{equation}
where
\begin{equation}
  J(\eta_1^\prime(x),\eta_2) =
  \frac{{\eta_1^\prime}^2-\frac{m_a^2\,m_b^2}{\eta_2^2\,S^2}}
       {2\,{\eta_1^\prime}^2 x}
   \left\{ \frac{{\eta_1^\prime}^2+\frac{m_a^2\,m_b^2}{\eta_2^2\,S^2}+\sqrt{\left(\frac{m_a^2\,m_b^2}{\eta_2^2\,S^2}
      +{\eta_1^\prime}^2\right)^2-4 \frac{m_a^2\,m_b^2}{\eta_2^2\,S^2}
      {\eta_1^\prime}^2 x^2}}{
    \sqrt{\left(\frac{m_a^2\,m_b^2}{\eta_2^2\,S^2}+{\eta_1^\prime}^2\right)^2-4
      \frac{m_a^2\,m_b^2}{\eta_2^2\,S^2}{\eta_1^\prime}^2 x^2}}\right\}\,.
\end{equation}
Note that the Jacobean reduces to the usual $1/x$ factor in the massless
limit.

After reversing the integration order, and performing the change of variable,
we find
\begin{eqnarray}
  \label{eq:inte_changed}
  \lefteqn{\int \done\eta_1\,\done\eta_2\,f_a(\eta_1)\,f_b(\eta_2)
    \int\limits_\alpha^1\done x\,K^{a,b}(x)
    \frac{\phi(x\,s_{ab})}{x\,\phi(s_{ab})} \done\PS{N} \,\Born_N(x\,s_{ab})
  }\nonumber\\
  &=&
  \int \done\eta_1^\prime\,\done\eta_2\, f_a(\eta_1^\prime)\,f_b(\eta_2)
  \done\PS{N} \,\Born_N(s_{ab})
  \int\limits_{\bar\alpha}^1\done x\,K^{a,b}(x)\left[J(\eta_1,\eta_2)
    \frac{\phi(s_{ab})}{x\,\phi(s_{ab}(\eta_1))}
    \frac{f_a(\eta_1)}{f_a(\eta_1^\prime)} \right] \, 
\end{eqnarray}
where $\eta_1 = \eta_1(\eta_1^\prime,x)$ is the old variable expressed in terms
of the new ones, and where the integration boundary is now given by
\begin{equation}
  \bar\alpha = \max\left\{\alpha,\eta_1^\prime\right\}
\end{equation}
Renaming $\eta_1 \leftrightarrow \eta_1^\prime$ finally yields
\begin{eqnarray}
  \label{eq:fin_I}
  \mathcal{I} &=&
  \frac{\alpha_s\, C_F}{2\,\pi}\,\int \done\eta_1\,\done\eta_2\,
  f_a(\eta_1)\,f_b(\eta_2)\,\done\PS{N} \,\Born_N(s_{ab})\nonumber\\
  &&\,\times\,\left\{\Virtual_N^{a,b}(\varepsilon) +
  \int\limits_{\bar\alpha}^1\done x\,K^{a,b}(x)\,
  \left[ J(\eta_1^\prime,\eta_2)\,
    \frac{\phi(s_{ab})}{x\,\phi(s_{ab}(\eta_1^\prime))}
    \frac{f_a(\eta_1^\prime)}{f_a(\eta_1)} -  1\right]
  - \int\limits_{0}^{\bar\alpha}\done x\,K^{a,b}(x)
  \right\}\,.
\end{eqnarray}
This disentangles the evaluation of the Born cross section from the $x$-integral such
that the whole curly bracket in Eq.~(\ref{eq:fin_I}) acts as a local $K$-factor
on top of the partonic cross section.

\subsection{Dipole formulae for initial-final configurations}
A detailed derivation of dipole formulae in the initial-final and final-initial
cases can be found in~\cite{Kotko:2012ui}, although in a slightly different
notation compared to the one presented in this work. In principle one could
also extract all relevant formulae from~\cite{Dittmaier:1999mb}
with the modifications described in the introduction, following the steps
presented in the previous section:

We consider the splitting: $Q_a\rightarrow g_k Q$ with spectator $i$ in the
final state.  To make the reading of this section more transparent, we
also report some useful kinematical quantities used throughout:
\begin{eqnarray}
  \begin{array}{lclclcl}
    Q^2   & = & (p_i \,-\,p_a\,+\,p_k)^2 ; &&
    \\[5mm]
    x_{ai} & = & \displaystyle
    \frac{p_a\cdot p_i \,+\,p_a\cdot p_k \,-\,p_i\cdot p_k}
         {p_a\cdot p_i + p_a\cdot p_k}\,; &&
    y_{a}  & = & \displaystyle
    \frac{p_a\cdot p_i }
         {p_a\cdot p_i + p_a\cdot p_k}\,;
    \\[5mm]
    R_{ai}(x) & = &\displaystyle
    \sqrt{\frac{\left(Q^2 + 2\,m_a^2\,x\right)^2 - 4\,m_a^2\,Q^2\,x^2}{\lambda_{ai}}}\,;&&
    \lambda_{ai} & = & \lambda(Q^2,m_a^2,m_i^2)\,.
  \end{array}
\end{eqnarray}

The subtraction term in this case is given by
\begin{equation}
  \IFDipole{a}{i}{k}(p_1,\dots, p_i,\dots,p_k,\dots,p_{N+1};p_a,p_b) =
  -\frac{ 1}{2\,x_{ai}\,p_a\cdot p_k}\frac{{\bf T}_a\cdot{\bf T}_i}{{\bf T}_a^2}
  \,{\bf V}^{ak}_i(p_a,p_i,p_k) 
  \otimes {\left|\widetilde{\mathcal{M}}_N
    \left(\widetilde{p}_1,\dots ,\widetilde{p}_{N+1};
    \widetilde{p}_a,p_b\right)\right|}^2,
\end{equation}
where the only divergent dipole contribution in the massive case reads
\begin{equation}
  \label{eq:if_dipoles}
  {\bf V}^{Q_ag_k}_i(p_a,p_i,p_k)\,=\,8\,\pi\,\mu^{2\varepsilon}\alpha_s\,C_F\,
  \left[\frac{2}{2-x_{ai}-z_{ai}}\,-R_{ai}(x_{ai})(1\,+\,x_{ai})-\varepsilon(1\,-\,x_{ab})-
    \frac{x_{ai}\,m_a^2}{p_a\cdot p_k}\right]\,.
\end{equation}

The mapped momenta can be expressed in terms of the original momenta using
\begin{eqnarray}
  \label{eq:trans}
  \widetilde{p}_i^\mu & = &
  \sqrt{\frac{\lambda_{ai}}{\lambda\left((p_i+k)^2,Q^2,m_a^2\right)}}\,
  \left(p_a^\mu-\frac{Q\cdot p_a}{Q^2}Q^\mu\right)\,+\,
  \frac{Q^2-m_a^2+m_i^2}{2\,Q^2}Q^\mu \,;\nonumber\\
  \widetilde{p}_a^\mu & = & \widetilde{p}_i^\mu - Q^\mu\,.
\end{eqnarray}

The integral of the extra emission phase-space can be split into two contributions,
as done in Eq.(\ref{eq:split_delta_k_term}), $\Virtual_{i,N}^{a}(\varepsilon)$ and
$\left[ K^{a}_i(x)\right]_+$.  They are given by
\begin{eqnarray}
  \label{eq:intsub}
  \Virtual_{i,N}^{a}(\varepsilon) &=&\;\;\displaystyle
  \frac{1}{\varepsilon}\,
  \left(1 + \frac{Q^2}{\sqrt{\lambda_{ai}}}\log\frac{c_1}{c_0}\right)
  + \log \left(\frac{\mu^2 m_i^2}{Q^4}\right)+\frac{1}{2}\nonumber\\
  && \displaystyle +\,
  \frac{Q^2} {\sqrt{\lambda_{ai}}}\left[\log\left(\frac{c_1}{c_0}\right) \log
   \left(\frac{m_a^2 \mu^2}{b_0^2 \lambda_{ai}}\right)-\log(c_1) \log
   \left(\frac{m_a^2+m_i^2-Q^2}{m_a^2}\right)\vphantom{\sum_{k=0}^{n}}\right. \nonumber \\
   && \hspace{3cm}+\left.\frac{1}{2} \log
     \left(\frac{c_1}{c_0}\right) \log (c_0 c_1)-\frac{\log(c_0)}{2}
     -2\sum_{k=0}^5(-1)^k {\rm Li}_2(c_k)\right]
  \nonumber \\
  && \displaystyle +\,
  \frac{Q^4\left(2 m_i^2+3 Q^2\right)}{2\sqrt{\lambda_{ai}}\left(m_i^2+Q^2\right)^2}
  \left[\log
    \left(\frac{\sqrt{\lambda_{ai}}-2 m_a^2-2 \gamma^2 Q^2+Q^2}{2 \gamma^2
      m_i^2}\right)+\frac{1}{\gamma}
    \log \left(\frac{\gamma  \left(\sqrt{\lambda_{ai}}+\gamma 
          Q^2\right)+2m_a^2}{(\gamma -1) \gamma  Q^2}\right) \right. \nonumber \\
       && \displaystyle \hspace{8.5cm}
        +\left. \frac{\left(\sqrt{\lambda_{ai}}+Q^2\right)
      \left(m_i^2+Q^2\right)}{2 m_a^2\left(2 m_i^2+3 Q^2\right)}\right]
\end{eqnarray}
where
\begin{eqnarray}
  \begin{array}{lclclcl}
    c_0   & = & \displaystyle
    \frac{Q^2+\sqrt{\lambda_{ai}}}{Q^2-\sqrt{\lambda_{ai}}} ; &&
    c_{1}  & = & \displaystyle
    \frac{Q^2-2\,m_i^2-\sqrt{\lambda_{ai}}}
         {Q^2-2\,m_i^2-\sqrt{\lambda_{ai}}}\,;
    \\[5mm]
    c_{2,3}   & = & \displaystyle
    \frac{\pm \sqrt{\lambda_{ai}}-  Q^2-2\,m_a^2}{2\,m_a^2} ; &&
    c_{4,5}  & = & \displaystyle
    2\,b_0\,\frac{m_i^2-m_a^2\mp \sqrt{\lambda_{ai}}}
                 {2\,m_a^2-Q^2 \mp \sqrt{\lambda_{ai}}}\,;
    \\[5mm]
    b_0 & = &\displaystyle
    \frac{2\,m_a^2}
         {\sqrt{\lambda_{ai}}-Q^2-2\,m_a^2}\,, &&
    \gamma & = &\displaystyle
    \frac{m_a^2}
         {\sqrt{-Q^2 -m_i^2 +i\epsilon}}\,;
  \end{array}
\end{eqnarray}
and
\begin{eqnarray}
  K_i^{a}(x) &=& -\frac{Q^2}{\sqrt{\lambda_{ai}}}\frac{1}{R_{ai}(x)}\left\{ \frac{2}{1-x}
  \log\frac{\left[1-z_1(x)\right]\left[2-x-z_2(x)\right]}
           {\left[1-z_2(x)\right]\left[2-x-z_1(x)\right]}
  +R_{ai}(x)(1+x)\log\frac{1-z_2(x)}{1-z_1(x)}\right.\nonumber \\
  && \left.
  +\frac{2\,m_a^2\,x^2}{Q^2}\left[\frac{1}{1-z_2(x)} -\frac{1}{1-z_1(x)}\right] \right\}
\end{eqnarray}
with
\begin{equation}
  z_{1,2}(x) = \frac{\left[Q^2-x\left(Q^2+2\,m_i^2\right)\right]
    \mp R_{ai}(x)\,(1-x)^2\sqrt{\lambda_{ai}}}
  {2\,\left[Q^2-x\left(Q^2-m_a^2\right)\right]}\,.
\end{equation}
The rest of the derivation follows exactly as in the previous section.

\section{\MCatNLO matching}
\label{sec:mcatnlo}
Having successfully built fixed-order NLO matrix elements in the 5FMS, we now
proceed to the matching to the parton shower along the lines of the
well-established \MCatNLO technique~\cite{Frixione:2002ik} as implemented
with small variations in the \Sherpa Monte Carlo~\cite{Gleisberg:2008ta},
and referred to as S-\MCatNLO~\cite{Hoeche:2011fd,Gehrmann:2012yg,Hoeche:2012yf}.

Note that our implementation closely follows that
of~\cite{Hoeche:2011fd,Gehrmann:2012yg}, which we refer to, for further details.
We start, by constructing the NLO-weighted Born cross section,
\begin{equation}
  \overline{\Born}(\PS{N}) =
  \Born(\PS{N})\,+\,\Virtual(\PS{N})\,+
  \,\mathcal{I}(\PS{N})
  \,+\,\int {\rm d} \Phi_{1}
  \left[D^{(A)}(\PS{N+1})\,\Theta(\mu_Q^2-t(\PS{N+1}))-D^{(S)}(\PS{N+1})\right]\,,
\end{equation}
where we have defined $D^{(A)}(\PS{N+1})$ such that, $\Real_N(\PS{N+1})$
can be split into an unresolved, divergent part, and a hard, resolved part,
\begin{equation}
  \Real_N(\PS{N+1}) = \,D^{(A)}(\PS{N+1})+\,\mathcal{H}(\PS{N+1})\,,
\end{equation}
and redefined Eq.~(\ref{eq:dds} )$\mathcal{S} \equiv D^{(S)}$ to be consistent
with the notation commonly used in this context.
It is worth mentioning at this stage that $D^{(A),(S)}$ have the same formal structure
and they only differ by finite terms\footnote{
  In their implementation in \Sherpa, in particular, they are equal up to phase-space,
  so that $D^{(A)}(\PS{N+1})=D^{(S)}(\PS{N+1})\Theta(\mu_Q^2-\kt^2)$}.
As a consequence they can both be written using the structure of
Eqs.~(\ref{eq:dds},\ref{eq:ddipoles}),
\begin{equation}
 D^{(A),(S)} \equiv \sum_{i=\text{FF,FI,IF,II}}\,\Dipole^{(A),(S)}_{i}\,.
\end{equation}

The last ingredient needed is the \MCatNLO Sudakov form factor.
This is built starting from $D^{(A)}(\PS{N+1})$,
\begin{equation}
  \Delta^{(A)}(t,t^\prime) = \exp\left\{-\int_t^{t^\prime}{\rm d}\phi_1\frac{D^{(A)}(\PS{N+1})}{\Born(\PS{N})}\right\}\,. 
\end{equation}
In particular, Eq.~(\ref{eq:dds}) implies that the Sudakov form
factor can be decomposed as
\begin{equation}
  \Delta^{(A)}(t,t^\prime) = 
  \exp\left\{-\sum_{i=\text{FF,FI,IF,II}}\int_t^{t^\prime}{\rm d}\phi_1\frac{D_{i}^{(A)}(\PS{N+1})}{\Born(\PS{N})}\right\}
  \equiv \prod_{i=\text{FF,FI,IF,II}}\Delta_i(t,t^\prime).
\end{equation}
The inclusion of mass effects in the initial state only
modifies the $i=II,IF$ contributions to $\Delta^{(A)}$ with respect
to their original definitions, which is what we focus on in the rest of this
section.

Finally, the \MCatNLO matched fully differential cross section can be written
in terms of the previous ingredients as
\begin{multline}
  \label{eq:mcatnlo}
  {\rm d}{\sigma}^{\text{\MCatNLO}}
  = {\rm d}\PS{N}\,\overline{\Born}(\PS{N})\,
  \left[\Delta^{(A)}(t_0,\mu_Q^2)\,+\,
    \int_{t\in [t_0,\mu_Q^2]}{\rm d}\PS{1}\,
    \frac{D^{(A)}(\PS{N+1})}{\Born(\PS{N})}\,\Delta^{(A)}(t(\PS{1}),\mu_Q^2)\right]\,F_J^{(N)}
  \\ +{\rm d}\PS{N+1}\, \mathcal{H}(\PS{N+1})\,F_J^{(N+1)}\,.
\end{multline}

\subsection{Sudakov form factor}
We now describe explicitly how the $\Delta_i$ contributions are constructed in our
implementation.  As already noted, only the $i=II,IF$ are changed with respect to
their original implementation, so in the following we restrict our discussion to them.
Most of the ingredients relevant to the matching can be obtained as the four-dimensional
limit -- $\varepsilon\rightarrow 0$ -- of the equations presented in Sec.\ref{sec:sub}.

One comment is in order here.  The initial state evolution is partially driven by
ratios of PDF factors at different scales, and possibly for different flavours for
transitions of quarks to gluons.  This may lead to a situation where such a factor
reads $f_g/f_Q$, {\it i.e.} the ratio of a gluon and a heavy quark PDF.  Ignoring effects
of intrinsic charm and beauty, the quark PDF has no support below its mass threshold,
and this ratio becomes ill defined.  Different solutions have been constructed in
various parton shower algorithms, most of which effectively enforce a splitting such that
the heavy quark is replaced by a gluon at threshold.

\subsubsection{Initial-Initial configurations}
First, we need to express the transverse momentum of the emission -- the ordering variable
in \Sherpa's parton showering -- in terms of the variables used to construct the subtraction,
$x$ and $y$:
\begin{equation}
  \kt^2 =
  \frac{2\, y\,\,(1-x-y)\,p_a\cdot p_b - (1-x-y)^2 m_a^2 - y^2 m_b^2}
       {1\,-\,\frac{m_a^2\,m_b^2}{(p_a\cdot p_b)^2}}.
\end{equation}
Further, we need the relevant Jacobian factor for the one particle
phase-space integration in~Eq.(\ref{eq:phase-space}),
\begin{equation}
  \frac{{\rm d}\kt^2}{\kt^2} =
  \frac{1-x-2\,y + (1-x-y)\frac{m_a^2}{p_a\cdot p_b} - y\frac{m_b^2}{p_a\cdot p_b}}
  {1-x-y - \frac{(1-x-y)^2}{2\,y}\frac{m_a^2}{p_a\cdot p_b}- \frac{y}{2}\,\frac{m_b^2}{p_a\cdot p_b}}\,
  \frac{{\rm d}y}{y}\,.
\end{equation}
The fully massive Sudakov form factor for initial-initial
dipole configurations is thus given by
\begin{equation}
  \Delta_{II}({\bf k}_{\perp,max}^2,{\bf k}_{\perp,0}^2) =
  \exp{\left\{ - \sum_{ak}\sum_{b\neq ak}\frac{1}{\mathcal{N}_{spec}}
    \int_{{\bf k}_{\perp,0}^2}^{{\bf k}_{\perp,max}}\frac{{\rm d}\kt^2}{\kt^2}
  \int_{x_{-}}^{x_{+}}{\rm d}x\,\mathcal{J}_{II}(x,y;\kt^2)\,{\bf V}^{ak,b}(p_a,p_b,p_k)\right\}}
\end{equation}
where
\begin{equation}
  \label{eq:jii}
  \mathcal{J}_{II}(x,y;\kt^2) =
  \frac{1-x-y - \frac{(1-x-y)^2}{2\,y}\frac{m_a^2}{p_a\cdot p_b}-
    \frac{y}{2}\,\frac{m_b^2}{p_a\cdot p_b}}
  {1-x-2\,y + (1-x-y)\frac{m_a^2}{p_a\cdot p_b} - y\frac{m_b^2}{p_a\cdot p_b}}\,
  \frac{s_{ab}}{\sqrt{\lambda_{ab}}}\,
  \frac{1}{x}\,\frac{f_a(\eta/x)}{f_a(\eta)}
\end{equation}
and ${\bf V}^{ak,b}$ can be taken from~Eq.(\ref{eq:diff_sub}).

\subsubsection{Initial-Final configurations}
Similarly to the previous case we get:
\begin{equation}
  \kt^2 =
  \frac{2\,p_a\cdot p_i (1-y) (1-x y) - m_i^2 (1-y)^2 -m_a^2 (1-x y)^2}
       {y^2 \left(1-\frac{m_a^2 m_a^2}{(p_a\cdot p_i)^2}\right)}.
\end{equation}
This implies that the Jacobian factor becomes
\begin{equation}
  \label{eq:jif}
  \mathcal{J}_{IF}(x,y;\kt^2) =
  \frac{2(1-y)(1-xy) - (1-y)^2\frac{m_i^2}{p_a\cdot p_i} - (1- xy)^2\frac{m_a^2}{p_a\cdot p_i}}
  {2- y- xy -(1-y)\frac{m_i^2}{p_a\cdot p_i} - (1- xy)\frac{m_a^2}{p_a\cdot p_i}}\,
  \frac{Q^2}{\sqrt{\lambda_{ai}}}\,
  \frac{1}{x}\,\frac{f_a(\eta/x)}{f_a(\eta)}\,,
\end{equation}
which in turn gives the fully massive initial-final contribution
to the Sudakov form factor
\begin{equation}
  \Delta_{IF}({\bf k}_{\perp,max}^2,{\bf k}_{\perp,0}^2) =
  \exp{\left\{ - \sum_{ak}\sum_{b\neq ak}\frac{1}{\mathcal{N}_{spec}}
    \int_{{\bf k}_{\perp,0}^2}^{{\bf k}_{\perp,max}}\frac{{\rm d}\kt^2}{\kt^2}
  \int_{x_{-}}^{x_{+}}{\rm d}x\,\mathcal{J}_{IF}(x,y;\kt^2)\,{\bf V}^{ak}_i(p_a,p_i,p_k)\right\}}
\end{equation}
where ${\bf V}^{ak}_i(p_a,p_i,p_k)$ is defined in Eq.~(\ref{eq:if_dipoles}).

\subsection{Physical kinematics}
In the practical implementation of the parton shower procedure, the extra
emission is attached to an underlying $N$ particle phase space. This
corresponds to the reversed procedure used to construct the reduced matrix
elements in Sec.\ref{sec:sub}.  The new momenta are then obtained from the
old ones by inverting~Eqs.(\ref{eq:trans}),
\subsubsection{Initial-Initial configurations}
For initial-initial configurations we get:
\begin{eqnarray}
  p_a^\mu &=& \frac{1}{x}\left[\sqrt{
      \frac{\left(2\,\widetilde{p}_a\cdot \widetilde{p}_b\right)^2-
        4\,m_a^2\,m_b^2\,x^2}
           {\lambda\left((\widetilde{p}_a+ \widetilde{p}_b)^2,m_a^2,m_b^2
             \right)}}
    \left(\widetilde{p}_a^\mu - \frac{\widetilde{p}_a\cdot \widetilde{p}_b}
         {m_b^2}\widetilde{p}_b^\mu\right)
         + \frac{\widetilde{p}_a\cdot \widetilde{p}_b}{m_b^2}
         \widetilde{p}_b^\mu\right]\nonumber\\
  p_b^\mu &=&  \widetilde{p}_b^\mu\nonumber\\
  p_k^\mu &=&
  \frac{1 -x -y -y\,\frac{m_b^2}{p_a\cdot p_b}}
       {1\,-\,\frac{m_a^2\,m_b^2}{(p_a\cdot p_b)^2}}\,p_a^\mu \,
       +\, \frac{y -(1-x-y)\,
         \frac{m_a^2\,x}{p_a\cdot p_b}}
       {1\,-\,\frac{m_a^2\,m_b^2}{(p_a\cdot p_b)^2}}\,p_b^\mu
       \,+\,\kt^\mu\,.
\end{eqnarray}
\subsubsection{Initial-Final configurations}
Similarly, for initial-final configurations,
\begin{equation}
  p_k^\mu =
  \frac{1}{y}\frac{1 -x y -(1-y)\,\frac{m_i^2}{p_a\cdot p_i}}
       {1\,-\,\frac{m_a^2\,m_i^2}{(p_a\cdot p_i)^2}}\,p_a^\mu \,
       +\, \frac{1}{y}\frac{1-y -(1-xy)\,
         \frac{m_a^2\,x}{p_a\cdot p_i}}
       {1\,-\,\frac{m_a^2\,m_b^2}{(p_a\cdot p_i)^2}}\,p_i^\mu
       \,+\,\kt^\mu\,.
\end{equation}
All other configurations are left unchanged by our scheme.

\section{Results}
\label{sec:results}
To study the impact of the inclusion of finite-mass effects, 
we compare results of the five-flavour massive scheme (5FMS) 
with the vanilla five flavour scheme, where $b$-quarks are massless.
In particular, just as an example and with no intentions of making any
statement about any beyond the Standard Model model, we focus on the production
of a scalar particle, $A$, coupling to $b$ quarks through a Yukawa coupling.
As a proxy to test the impact of the inclusion of mass effects we vary the mass of this
scalar particle in the range $20-500$~GeV. Further, we let the coupling of
the $b$-quarks to this particle vary too by varying the parameter
$\tan\beta$, mimicking a two Higgs doublet model.

\begin{figure}[!h]
  \begin{center}
    \includegraphics[scale=0.7]{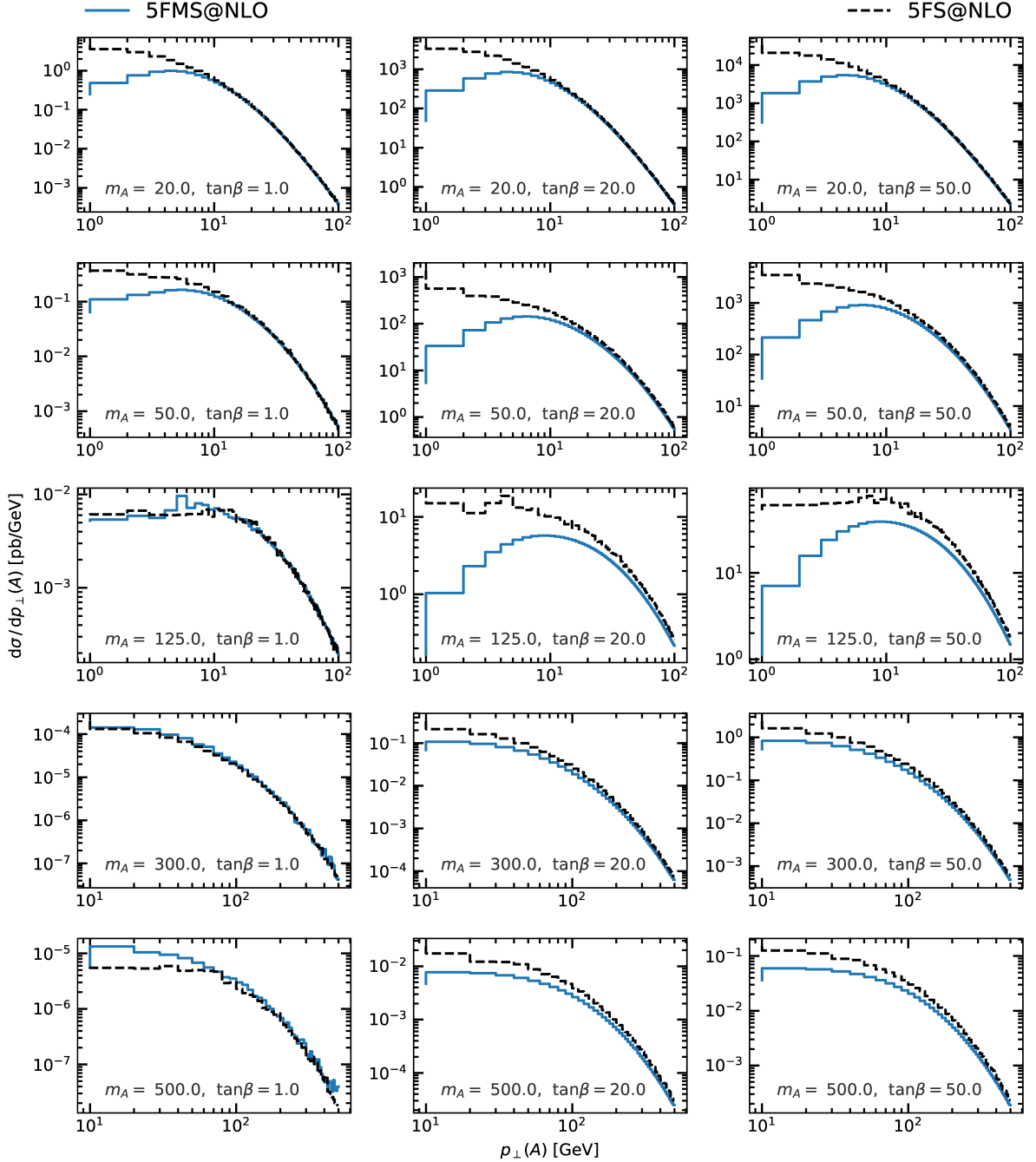}
    \caption{The $p_\perp$ spectrum of the scalar boson $A$ for
      various combinations of $m_A$ and $\tan\beta$, in the 5FS and in the 5FMS, fixed-order
      predictions.}
    \label{fig:matb_fo}
  \end{center}
\end{figure}

\begin{figure}[!h]
  \begin{center}
    \includegraphics[scale=0.7]{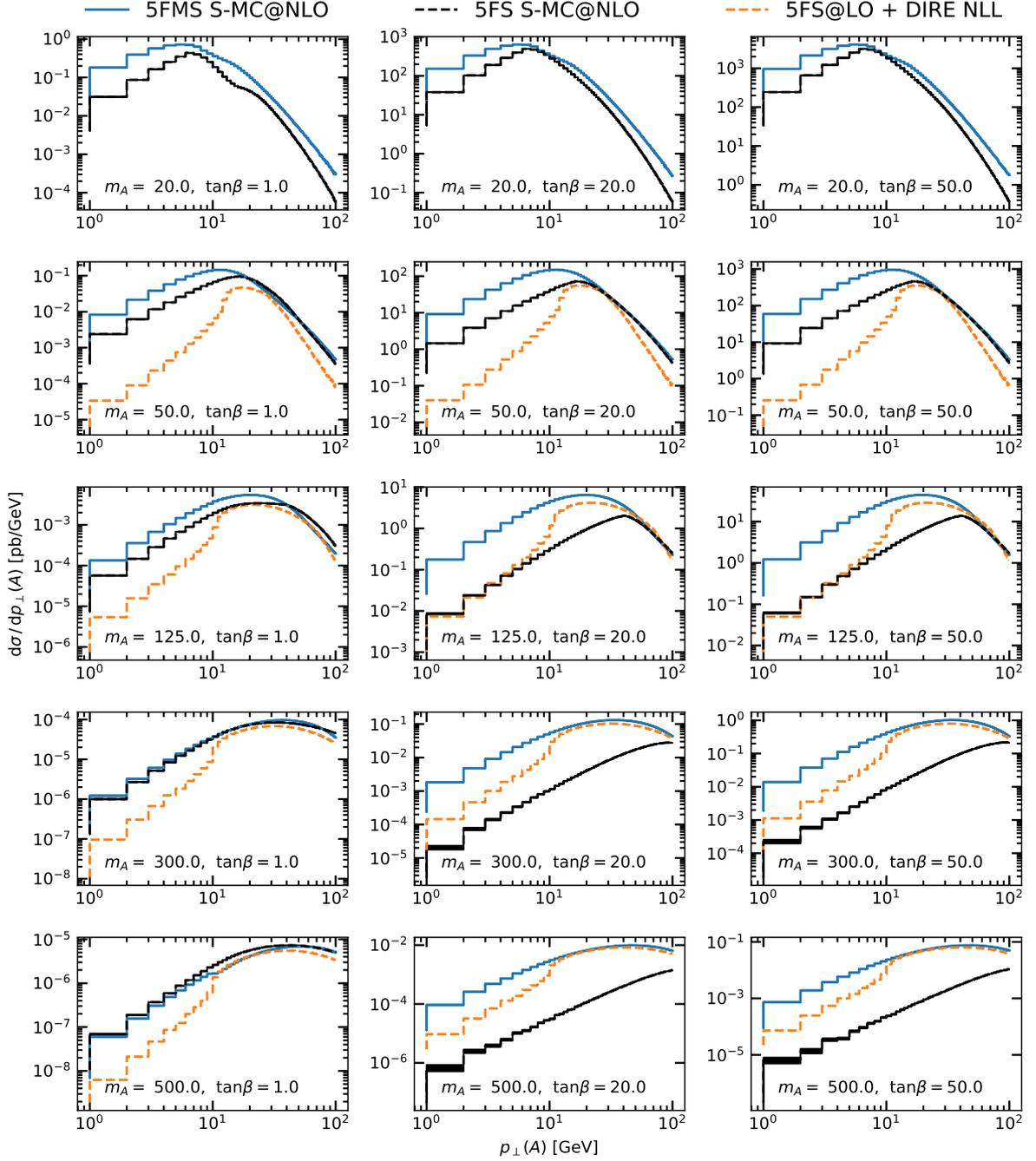}
    \caption{The $p_\perp$ spectrum of the scalar boson $A$ for
      various combinations of $m_A$ and $\tan\beta$. The 5FS and the 5FMS are
      computed at \MCatNLO accuracy while the \Dire prediction is obtained
      matching a LO matrix element with the NLL
      shower~\cite{Hoche:2017iem,Hoche:2017hno}.}
    \label{fig:matb_ps}
  \end{center}
\end{figure}

We study this process both at fixed-order next-to-leading order accuracy,
and at \MCatNLO accuracy.
Plots for the former are collected in Fig.~(\ref{fig:matb_fo}),
while the latter set-up is shown in Fig.~(\ref{fig:matb_ps}).
All our results are obtained within the \Sherpa event
generator~\cite{Gleisberg:2008ta}.  Leading-order matrix elements,
including those of real radiation processes, are calculated using the
\Amegic~\cite{Krauss:2001iv} matrix element generator.  The differential
subtraction follows closely the algorithms of~\cite{Gleisberg:2007md}, extended
with the ingredients reported in the previous Section, Sec.~\ref{sec:mass_sub}.
The integrated subtraction terms are implemented in \Sherpa and will be made
publicly available in a future \Sherpa release.  Virtual corrections have
been obtained from the \OpenLoops generator~\cite{Cascioli:2011va}.
Finally, for the \MCatNLO results, we
make use of the modifications described in Sec.~\ref{sec:mcatnlo}
to \Sherpa standard parton shower, the \CSS shower~\cite{Schumann:2007mg}.

We apply no cuts at generation level.  However, in the following we define as
$b$-jet any jet with $p_\perp\geq 25$~GeV that has at least one $b$-flavoured parton
in it. We further require any particle in the final state to
have $|\eta|\leq 2.5$. Jets are clustered using the
anti-$k_T$~\cite{Cacciari:2008gp} algorithm as implemented in
\Fastjet~\cite{Cacciari:2011ma}
while the event analysis is performed using the \Rivet
package~\cite{Waugh:2006ip,Buckley:2010ar}.
We set the renormalization, factorisation and shower starting
scales to be $\mu_R=\mu_F=\mu_Q=m_A/3$. 

For fixed-order predictions, in Fig.~(\ref{fig:matb_fo}),
we use the -- \Sherpa default -- NNPDF30 NNLO set with
$\alpha_s(m_Z)=0.118$~\cite{Ball:2014uwa} for the 5FS and the 5FMS.
Results presented in Fig.~(\ref{fig:matb_ps}), however, where we
further compare the two five-flavour scheme predictions with a LO matrix 
element matched with
the NLL \Dire parton shower~\cite{Hoche:2017iem,Hoche:2017hno},
we employ the CT14 NNLO PDF set~\cite{Dulat:2015mca},
to satisfy \Dire's requirement of positive definite PDFs, for all predictions.
Note that the \Dire prediction is also with five massless flavour, like
the standard 5FS.

We start our discussion from the fixed-order results, Fig.\ref{fig:matb_fo}.
In this case, we have two mass scales, $p_\perp$ and $m_A$. Further, as we let
them vary, the mass corrections differ in size in different regions of
phase-space.  In any case, we always expect the 5FS and the 5FMS to converge
to each other in the region $p_\perp\sim m_A$, which is indeed evident in
Fig.\ref{fig:matb_fo}.  Note the different $p_\perp$ range in Fig.\ref{fig:matb_fo}
for $m_A \geq 300$~GeV to show this expected behaviour.  As $m_A$ increases we
expect a reduction in the absolute size of mass effects, with differences
between the two schemes remaining in the region $p_\perp\sim m_b$. 

Looking at the standard $(m_H=125,\tan\beta=1)$ point, 
we expect mass effects to play a marginal role, of order $\sim 1-5\%$, at the
level of total cross sections. Furthermore, since they are power suppressed,
we expect them to be less important at large ${p}_{\perp}$, while having the
largest impact in the lower bins of the distribution.  This is due to the fact
that the difference in the mass treatment between the two schemes is only in the hard
matrix elements and phase-space.  This is confirmed in our plot 

We now turn to the parton shower-matched results, Fig.\ref{fig:matb_ps}.
Leaving for a second the discussion of the \Dire prediction aside, in general,
we expect a shower prediction to fall on the fixed order result for values
of $p_\perp \gtrsim \mu_Q$, where the discussion of fixed order results apply.
In this case, too, we expect less important mass effects in the region where
both $m_A$ and $p_\perp$ are large, but not for large $m_A$ and small $p_\perp$.
These general considerations are all confirmed in~Fig.\ref{fig:matb_ps}.
The additional sample is obtained using \Dire. As the implementation of this
shower stops the emission for scales $\sim 2\,m_b$, no transverse momentum
is generated for the scalar particle when $m_A=20$~GeV, with
$\mu_Q = m_A/3 < 2\,m_b$.  As for all other configurations the the 5FMS and
\Dire predictions are in good agreement, in the region of the Sudakov peak,
$p_\perp\lesssim \mu_Q$.

\section{Conclusions}
In this paper we presented all the ingredients
that are necessary to construct an extension of VFNS, like the 5FS,
in which heavy quarks are treated as massless partons, to allow for massive
partons in the initial state, for processes at hadron colliders, at NLO accuracy.
In particular we extended the successful Catani--Seymour scheme for
subtraction of infrared divergences to the case of massive initial states.  In
variance to an earlier treatment by Dittmaier, we do not use a finite photon
mass as regulator but consistently work in $D=4+2\varepsilon$ dimensions.  We
also re-parametrize the result for the integrated subtraction term in such a
way that the residual integral over the light-cone fraction of the emitted
particle decouples from the evaluation of the Born cross section, rendering
our result more useful for direct implementation.
Further, we used these massive dipoles to extend the S-\MCatNLO matching
as well as the shower generation.

We investigated the effect of finite quark masses at fixed-order accuracy in
the process $b\bar{b}\to A$.  Mass effects for this process are generally
quite small, both at the inclusive and differential level, of the order of a
few percent.  This however might not hold true for other processes involving
heavy quarks.  A five flavour massive scheme will provide insight by producing
fully differential results including mass effects in a consistent way at
matrix-element level.
As an additional example we presented simulations for the production
of a scalar particle $A$ in bottom-quark fusion, for various configurations
of $m_A$ and $\tan\beta$.

\section*{Acknowledgements}
We want to thank our colleagues from the \Sherpa collaboration for animated
discussions, technical and moral support.  We are particularly grateful to
Stefan H\"oche for his help in the implemention of the method and the running
of the \Dire parton shower.
This work has received funding from the European Union's Horizon 2020
research and innovation programme as part of the Marie Skłodowska-Curie
Innovative Training Networks ``MCnetITN3'' (grant agreement no. 722104)
and ``HiggsTools'' (grant agreements PITN-GA2012-316704) and by the ERC
Advanced Grant MC@NNLO (340983).
\bibliographystyle{amsunsrt_modp}
\bibliography{bbb}
\end{document}